\documentclass[fleqn,twoside]{article}
\usepackage{espcrc2}
\usepackage{graphicx}
\usepackage[figuresright]{rotating}

\newcommand{\AmS}{{\protect\the\textfont2
  A\kern-.1667em\lower.5ex\hbox{M}\kern-.125emS}}

\newcommand{\VEV}[3]{\left\langle #1\left| #2 \right| #3\right\rangle}
\newcommand{\vev}[1]{\left\langle #1 \right\rangle}

\newcommand{\ovl}[1]{\overline{#1}}

\newcommand{\re}{{\rm Re}}

\title{Kaon matrix elements in domain-wall QCD with DBW2 gauge action}

\author{J.~Noaki\address[RBRC]{RIKEN BNL Research Center, Brookhaven
National Laboratory, Upton, NY 11973-5000, USA}\ \  for RBC
Collabolation\thanks{
We thank RIKEN, BNL and the U.S.\ DOE for providing the facilities
essential for the completion of this work.}}

\begin{document}
\begin{abstract}
We present calculations of the decay constants and 
kaon B-parameter $B_K$ as the first stage of RBC Collaboration's 
quenched numerical simulations using DBW2 gauge action and domain-wall 
fermions. 
Some of potential systematic errors and consistency to previous works
are discussed. 
\vspace{1pc}
\end{abstract}
\maketitle

\section{Introduction}

In the quantities related to kaon physics such as
$f_\pi$, $f_K$, $B_K$, $\re A_0/\re A_2$ and $\epsilon'/\epsilon$,
the first three are constructed straightforwardly on the lattice. 
However, in actual numerical simulations, there are several
potential sources of systematic error, namely explicit breaking of 
chiral symmetry, scaling violation, finite volume effect and 
quenching effect. 
Although many efforts have been made to decrease these errors,  
yet more numerical simulations should be done to attain a conclusion.

We report our results of $f_\pi$, $f_K$ and $B_K$ as a conclusion of 
the first stage of our longstanding quenched numerical simulations 
using domain-wall fermion and DBW2 gauge action.
By our choice of the lattice action, it is expected that the 
contamination from the explicit chiral symmetry breaking is 
vanishingly small. In particular, we estimate the effect of operator mixing
in the calculation of $B_K$ through the non-perturbative calculation of the
renormalization factors which weight contributions of the mixing operators.
We also examine the scale dependence of our results by 
carrying out two kinds of numerical simulations with a similar 
physical lattice volume $\approx 1.6$ fm but with different 
scales $a^{-1}\approx$ 2 GeV and 3 GeV.
After comparing our results with those from previous works,
we present our continuum results of $B_K$.

For other kind of RBC's numerical simulations with the dynamical 
quark, see ref.~\cite{DYNAMICAL}.

\section{Numerical Simulations}
\begin{table}
\caption{Simulation parameters, statistics and results 
of $a^{-1}$,  $m_{\rm res}a$, $Z_A$ and $m_sa/2$.}
\label{PARAMS}
\begin{tabular}{rcc}
\hline
$\beta$      & $1.22$          & $1.04$          \\
size         & $24^3\times 48$ & $16^3\times 32$ \\
 $M_5a$      & $1.65$          & $1.8$           \\
 $L_s$       & $10$            & $16$            \\
 $m_fa$      & 0.008 -- 0.040  & 0.01 -- 0.05    \\
                           & in step of 0.008& in steps of 0.01\\
\hline\hline
\multicolumn{2}{l}{\#configs.} & \\
($Z_{B_K}$)& 53              & 50              \\
 (others)  & 106             & 202             \\

\hline\hline
\multicolumn{2}{l}{basic results }& \\
 $a^{-1}$     & 2.914(54) GeV   & 1.982(21) GeV   \\
$m_{\rm res}a$& $9.73(4)\cdot 10^{-5}$ & $1.85(12)\cdot 10^{-5}$\\
        $Z_A$ & 0.88813(19)     & 0.84019(17) \\
\hline
\end{tabular}
\vspace{-0.4cm}

\end{table}

In Table~\ref{PARAMS}, we enumerate two sets of our simulation 
parameters, statistics and results of the basic quantities such as 
lattice scale from the input $m_\rho=770$ MeV, the residual quark
mass $m_{\rm res}$ and the renormalization factor of axial vector $Z_A$.
Results of the basic quantities for $\beta=1.04$ are quoted from 
ref.~\cite{RBCDBW2}, in particular. 
For the finer lattice with $\beta=1.22$, it is known that the topological 
charge $Q_{\rm top}$ changes very slowly in the ordinary Markov 
chain~\cite{RBCDBW2}.
To avoid incorrect distribution of $Q_{\rm top}$,
we generated configurations as described in ref.~\cite{LAT02NOAKI}.
As a result, we obtained a reasonable distribution of the topological 
charge: $\vev{Q_{\rm top}}=0.38(29)$.
We employ the averaged quark propagator over those with
periodic and anti-periodic boundary conditions in the temporal direction 
for the calculation of $f_\pi$, $f_K$ and $B_K$. Therefore, temporal
lattice size may be treated as 96 and 48 for $\beta=1.22$ and $1.04$.

\section{Decay Constants}

\begin{figure}[t]
\begin{center}
\includegraphics[width=5.5cm,clip]{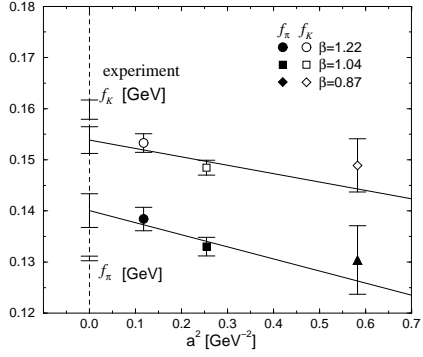}
\end{center}
\vspace*{-1.0cm}

\caption{Results of $f_\pi$ (filled symbols) and $f_K$ [GeV] (open symbols) 
versus $a^{2}\ [{\rm GeV}^{-2}]$. }
\label{FPI_COMP}
\vspace*{-0.6cm}

\end{figure}
As a combination of fit parameters of two point correlation functions
of mesons,
we calculate decay constants of the pseudo-scalar from the combination
of the fit parameters:
\begin{eqnarray}
f_{PS}= \frac{{\cal A}^{A_4P}_{\rm pw}}
{\sqrt{\frac{m_{\rm  PS}}{2}V {\cal A}^{PP}_{\rm ww}}},\label{type1}
\end{eqnarray}
where ${\cal A}^{A_4P}_{\rm pw}$ and ${\cal A}^{PP}_{\rm ww}$
are the amplitudes of the correlation functions 
$\vev{A^{\rm point}_4(t)P^{\rm wall}(0)}$ and 
$\vev{P^{\rm wall}(t)P^{\rm wall}(0)}$. $m_{PS}$ is determined from
the simultaneous fit of these two correlation functions thus common to both.
Using the linear fit with $f_{PS}= c_0 + c_1m_{PS}^2$, 
we extract bare values of $f_\pi$ at $m_{PS}=m_\pi$ and $f_K$ 
at $m_{PS}= m_K$. Multiplying these values by $Z_A$ in Table~\ref{PARAMS},
we show our physical results in Figure~\ref{FPI_COMP}. As well as for $\beta
=1.22$ (circles) and $1.04$ (squares), results for $\beta=0.87$ 
(diamonds) from ref.~\cite{RBCDBW2} are plotted in this figure.
By the linear fit using these three data, our results are 
$f_\pi=$ 140.1(3.3) MeV and $f_K= $153.9(2.6) MeV in the continuum limit.
They are both inconsistent with the experimental values.
On the other hand, our result $f_K/f_\pi= 1.098(13)$ in the continuum limit
is roughly consistent with the estimation from the quenched chiral 
perturbation theory~\cite{BernardGolterman}.

\section{Kaon B-parameter $B_K$}

\begin{figure}[t]
\hspace{-0.7cm}
\includegraphics[width=4.16cm,clip]{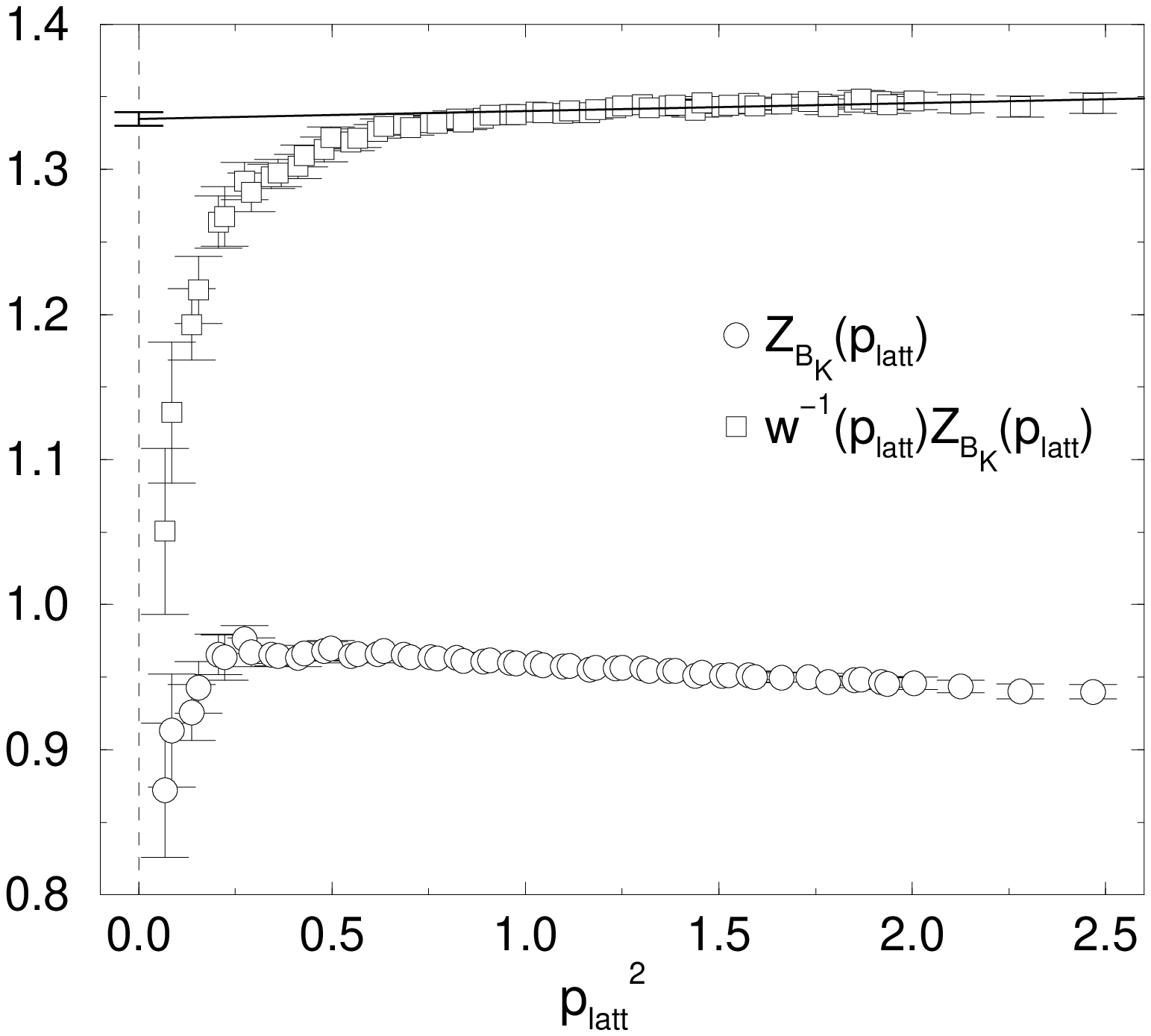}
\includegraphics[width=3.8cm,clip]{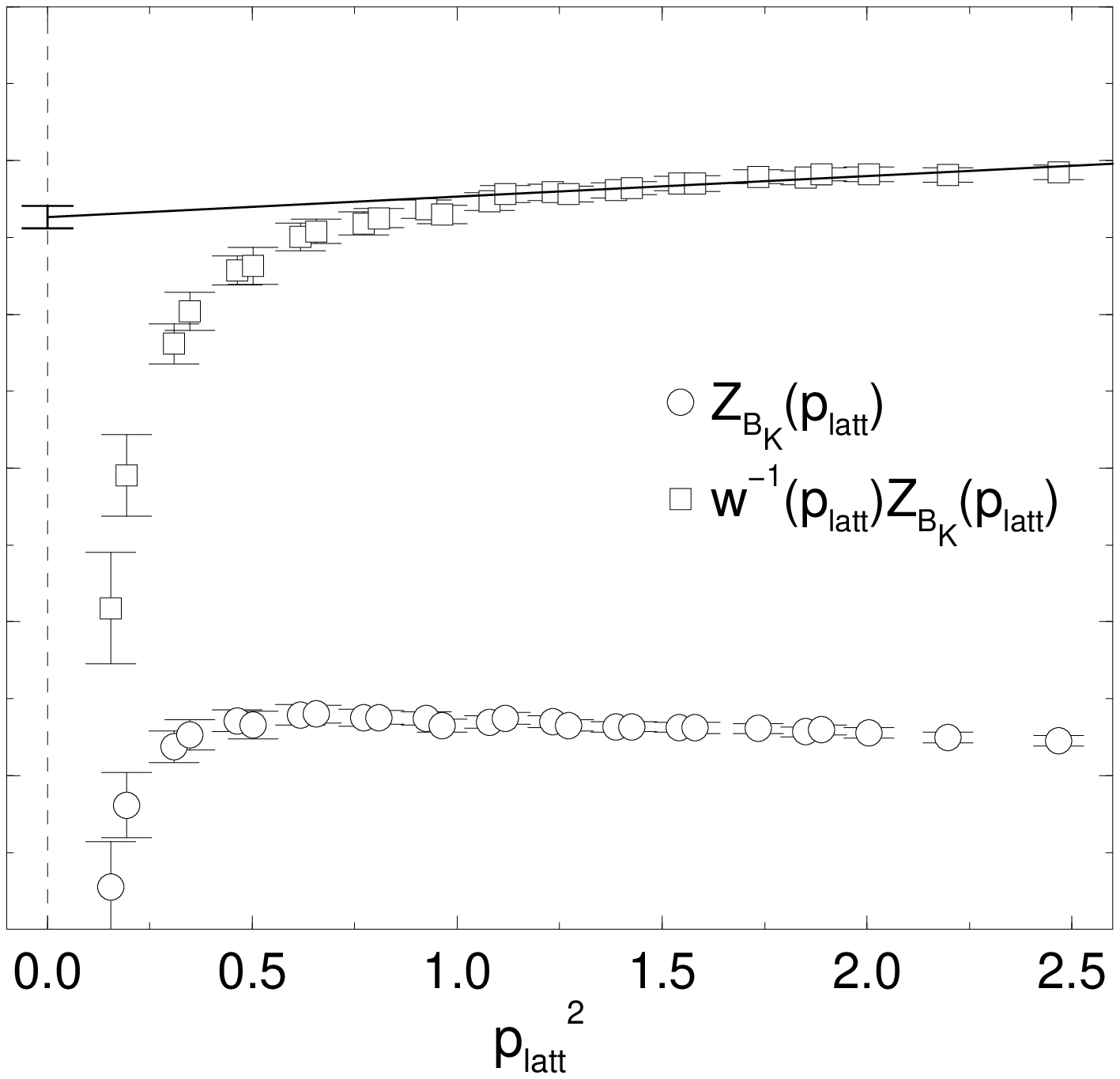}
\vspace*{-1.1cm}

\caption{ $Z_{B_K}$ and RGI value $\hat{Z}_{B_K}$ versus 
$p_{\rm latt}^2$ for DBW2 $\beta=1.22$ (left) and $\beta=1.04$ (right). 
In each panel, $Z_{B_K}$ (circles), RGI value 
(squares) and its linear extrapolation are shown.}
\label{ZBKinv}
\vspace*{-0.4cm}

\end{figure}
We compute $B_{PS}$ {\it i.e.} $B_K$ on the lattice, by extracting a
plateau of the ratio of the correlation functions:
\begin{eqnarray}
 \frac{\VEV{0}{P^{\rm wall} (t'_0) Q_{\Delta S=2}(t)P^{\rm wall}(t_0)}{0}}
      {\frac{8}{3} \VEV{0}{P^{\rm wall}(t'_0)A_4(t)}{0}
                   \VEV{0}{A_4(t)P^{\rm wall}(t_0)}{0}},
\end{eqnarray}
where the locations of two wall sources $P^{\rm wall}$ are 
$(t_0, t'_0) = (7, 41)$ and $(5, 27)$ for $\beta = 1.22$ and $1.04$.
We chose $19\le t\le 29$ and $14 \le t \le 17$ for the fit ranges in 
each case.

To determine the renromalization factor $Z_{B_K}$, 
we employ the non-perturbative calculation 
which was pioneered by ref.~\cite{NPR}.
Including the possibility of the contribution from the mixing operators,
the renormalization condition is written by
\begin{eqnarray}
Z_q^{-2}Z_{ij}\Gamma^{(4){\rm latt}}_{{\cal O}_j}(p_{\rm latt})
  = \Gamma^{(4){\rm tree}}_{{\cal O}_i},\label{renorm2}
\end{eqnarray}
where $\Gamma^{(4){\rm latt}}_{{\cal O}_j}$ and 
$\Gamma^{(4){\rm tree}}_{{\cal O}_j}$ are the amputated 
four point vertices on the lattice and in the tree level for 
the four-quark operator ${\cal O}_j$ with certain chiralities 
$i = VV\pm AA,\ SS\mp PP$ and $TT$. By solving (\ref{renorm2}), 
we observe that the renormalization factors $Z_{ij}$ for the mixing 
operators are at most $\sim 0.1\%$ of the factor for $VV+AA$, 
which corresponds to $Q_{\Delta S=2}$. These small $Z_{ij}$'s suppress 
the effect of the operator mixing to be less than the statistical error 
of $B_{PS}$ in spite of the fact the B-parameters of
the mixing operators are a few dozens times larger than $B_{PS}$~\cite{RBCbk}.

In Figure~\ref{ZBKinv}, we show the results 
$Z^{\rm RI/MOM}_{B_K}= Z_{VV+AA,VV+AA}/Z_A^2$ in RI/MOM scheme 
and the renormalization group independent (RGI) value 
$\hat{Z}_{B_K}= w^{-1}_{\rm RI/MOM}(p_{\rm latt})Z^{\rm RI/MOM}(p_{\rm latt})$,
where the factor $w^{-1}_{\rm RI/MOM}$ was calculated in ref.~\cite{Ciuchini}.
Extrapolating the data for $p_{\rm latt} >1$ linearly, we 
quote $\hat{Z}_{B_K}$ at $p_{\rm latt}=0$.
 Renormalization factor in the $\ovl{\rm MS}$, NDR scheme are calculated as 
$Z^{\ovl{\rm MS}}_{B_K}= w_{\ovl{\rm MS}}(\mu= 2\ {\rm GeV})\hat{Z}_{B_K}$.

\begin{figure}[t]
\begin{center}
\includegraphics[width=5.5cm,clip]{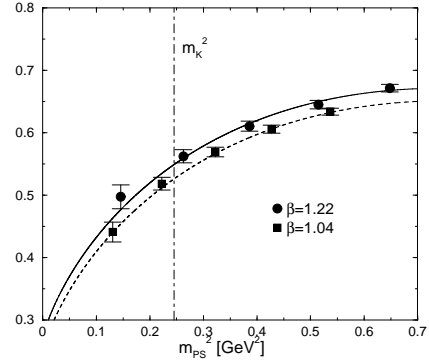}
\end{center}
\vspace*{-1.2cm}

\caption{$B^{\rm (ren)}_{PS}$ vs. 
$m_\pi^2\ [{\rm GeV}^2]$ with fit curves.}
\label{BKren_sum}
\vspace*{-0.4cm}

\end{figure}

\begin{figure}[t]
\begin{center}
\includegraphics[width=5.5cm,clip]{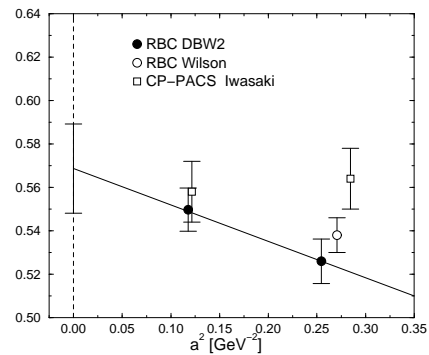}
\end{center}
\vspace*{-1.2cm}

 \caption{$B_K$ versus $a^{-2}\ [{\rm GeV}^{-2}]$. 
 As well as our results (filled symbols), we quoted previous results 
from refs.~\cite{CPPACSBK,RBCep} (open symbols).}
\label{BK}
\vspace*{-0.5cm}

\end{figure}

We plot renormalized value 
$B^{\rm (ren)}_{PS}=Z^{\ovl{\rm MS}}_{B_K}(\mu=2\ {\rm GeV})B_{PS}$ 
as a function of $m_{PS}^2\ [{\rm GeV}^2]$ in Figure~\ref{BKren_sum}, 
where circles are from $\beta = 1.22$ and squares from $1.04$. 
Using the chiral expansion~\cite{Sharpe} 
\begin{eqnarray}
B_{PS}= \xi_0\left[ 1- \frac{6}{(4\pi f)^2}m^2_{PS}\ln 
\frac{m_{PS}^2}{\Lambda_\chi^2} \right] + \xi_1 m_{PS}^2,
\label{BKratio_clog}
\end{eqnarray}
and our results of $f_{PS}(m_{PS}=0)$ for the coefficient of the chiral log,
we obtain the solid and dashed curves for $\beta=1.22$ and $1.04$ with
$\chi^2/$dof = 0.80 and 0.08, respectively.
Result of $B_K^{\ovl{\rm MS}}(\mu=2\ {\rm GeV})$ for each $\beta$ 
from the interpolation to $m_{PS}=m_K$ are shown by the filled symbols 
in Figure~\ref{BK}.
In the same figure, we plot results of CP-PACS~\cite{CPPACSBK} 
using Iwasaki gauge action with two set of similar parameters as ours,
{\it i.e.}
$(a^{-1} = 2.87\ {\rm GeV},\ 24^3\times 60,\ L_s=16)$ and
$(a^{-1} = 1.88\ {\rm GeV},\ 16^3\times 40,\ L_s=16)$
with the open squares. On the other hand, the open circle corresponds 
to the previous result of RBC~\cite{RBCep} using Wilson gauge action
with $(a^{-1} = 1.922(40)\ {\rm GeV},\ 16^3\times 32,\ L_s=16)$.
For $a^{-1} \approx$ 2 GeV, results from these three kinds of calculations 
distribute over the width of more than 10\%. On the other hand, for 
$a^{-1}\approx 3$ GeV, results from both Collaboration agree with each other.
In the continuum limit taken by the naive extrapolation of our two
results, we extract $B_K^{\ovl{\rm MS}}(\mu=2\ {\rm GeV})= 0.569(21)$
and $\hat{B}_K=0.762(28)$.

\newcommand{\NP}{Nucl.~Phys.}
\newcommand{\NPSup}{Nucl.~Phys.~{\bf B} (Proc.~Suppl.)}
\newcommand{\PL}{Phys.~Lett.}
\newcommand{\PRD}{Phys.~Rev. D}
\newcommand{\PRL}{Phys.~Rev.~Lett.}

\end{document}